\documentclass[pdflatex,sn-mathphys-num]{sn-jnl}

\usepackage{graphicx}%
\usepackage{multirow}%
\usepackage{amsmath,amssymb,amsfonts}%
\usepackage{amsthm}%
\usepackage{mathrsfs}%
\usepackage[title]{appendix}%
\usepackage{xcolor}%
\usepackage{textcomp}%
\usepackage{manyfoot}%
\usepackage{booktabs}%
\usepackage{algorithm}%
\usepackage{algorithmicx}%
\usepackage{algpseudocode}%
\usepackage{listings}%
\usepackage{enumitem,kantlipsum}

\usepackage{titlesec}

\titleclass{\subsubsubsection}{straight}[\subsubsection]
\newcounter{subsubsubsection}[subsubsection]
\renewcommand\thesubsubsubsection{\arabic{subsubsubsection})}
\titleformat{\subsubsubsection}
  {\normalfont\normalsize\bfseries\itshape}{\thesubsubsubsection}{1em}{\normalfont\normalsize\bfseries\itshape}
\titlespacing*{\subsubsubsection}
  {0pt}{3.25ex plus 1ex minus .2ex}{1.5ex plus .2ex}

\usepackage{tabularx}

\theoremstyle{thmstyleone}%
\theoremstyle{thmstyletwo}%

\theoremstyle{thmstylethree}%

\begin{document}

\title[Article Title]{MuDiT \& MuSiT: Alignment  with Colloquial Expression in Description-to-Song Generation}


\author[1,2]{\fnm{Zihao} \sur{Wang}}\email{carlwang@zju.edu.cn}

\author[1,2]{\fnm{Haoxuan} \sur{Liu}}\email{liuhaoxuan@zju.edu.cn}
\equalcont{These authors contributed equally to this work.}

\author[1,2]{\fnm{Jiaxing} \sur{Yu}}\email{yujx@zju.edu.cn}
\equalcont{These authors contributed equally to this work.}

\author[2]{\fnm{Tao} \sur{Zhang}}\email{zhangtao8@gmail.com}

\author[2]{\fnm{Yan} \sur{Liu}}\email{liuyan@liao.com}

\author*[1]{\fnm{Kejun} \sur{Zhang}}\email{zhangkejun@zju.edu.cn}

\affil*[1]{\orgdiv{College of Computer Science and Technology}, \orgname{Zhejiang University}, \orgaddress{\street{No.38 Zheda Road}, \city{Hangzhou}, \postcode{310027}, \state{Zhejiang}, \country{China}}}

\affil[2]{\orgdiv{Caichong Artificial Intelligence Research}, \orgname{DuiNiuTanQin Co., Ltd.}, \orgaddress{\street{No.10 Jiuxianqiao North Road}, \city{Beijing}, \postcode{100015}, \country{China}}}


\abstract {
    Amid the rising intersection of generative AI and human artistic processes, this study probes the critical yet less-explored terrain of alignment in human-centric automatic song composition.
    We propose a novel task of Colloquial Description-to-Song Generation, which focuses on aligning the generated content with colloquial human expressions. This task is aimed at bridging the gap between colloquial language understanding and auditory expression within an AI model, with the ultimate goal of creating songs that accurately satisfy human auditory expectations and structurally align with musical norms.
    Current datasets are limited due to their narrow descriptive scope, semantic gaps and inaccuracies. To overcome data scarcity in this domain, we present the Caichong Music Dataset (CaiMD). CaiMD is manually annotated by both professional musicians and amateurs, offering diverse perspectives and a comprehensive understanding of colloquial descriptions. Unlike existing datasets pre-set with expert annotations or auto-generated ones with inherent biases, CaiMD caters more sufficiently to our purpose of aligning AI-generated music with widespread user-desired results.
    Moreover, we propose an innovative single-stage framework called MuDiT/MuSiT for enabling effective human-machine alignment in song creation.  This framework not only achieves cross-modal comprehension between colloquial language and auditory music perceptions but also ensures generated songs align with user-desired results. MuDiT/MuSiT employs both a fine-tuned LLM for lyrics creation bestows it with proficiency in emulating standard song structures effectively, and one DiT/SiT model for end-to-end generation of other musical components like melody, harmony, rhythm, vocals, and instrumentation. The approach ensures harmonious sonic cohesiveness amongst all generated musical components, facilitating better resonance with human auditory expectations. 
    Experimental results showcase MuDiT/MuSiT's superior performance over open-source frameworks and demonstrate comparable generation quality to proprietary models alongside better alignment with colloquial song descriptions. 
    Hence, our research pioneers efficient, symbiotic human-AI alignment in automatic song composition, enhancing the collaborative experience between humans and AI in artistic endeavors.
}

\keywords{Description-to-song generation, AI alignment, Colloquial expression, Data scarcity, Cross-modal comprehension, Audio quality evaluation}



\maketitle

\section{Introduction}

Songs are an integral part of human culture, functioning as a universal medium for communication. The creation of music necessitates significant time and effort from composers to attain the anticipated results. Recently, the swift development of datasets~\cite{bertin2011million,yang2017midinet,bogdanov2019mtg,wang2020pop909,wang2022songdriver,huang2023noise2music,melechovsky2023mustango,lu2023musecoco,schneider2023mo,zhu2023ernie} and generative models~\cite{sheng2021songmass,zeng2021musicbert,hsiao2021compound,yu2022museformer,agostinelli2023musiclm,wu2023melodyglm,schneider2023mo,huang2023noise2music,copet2024simple} has catalyzed rapid advances in automatic song composition. Nonetheless, as far as we are aware, at a time when the alignment efforts between generative AI in the arts and human are becoming increasingly crucial, the alignment work in automatic song composition AI has not received adequate attention. 
AI alignment~\footnote{https://en.wikipedia.org/wiki/AI\_alignment} typically refers to ensuring that AI systems can maintain consistency with human values during their operation, which is a concern highly regarded by researchers such as Geoffrey Hinton and Ilya Sutskever (former Chief Scientist of OpenAI~\footnote{https://openai.com}). 
In the context of automatic song composition tasks, this translates into whether AI models can fully comprehend human colloquial descriptions, whether the generated songs meet the auditory expectations desired by humans, and whether they conform to human anticipations for song structure.
`Humans' here does not exclusively pertain to musical professionals but also encompasses a vast number of amateurs. This implies that beyond understanding professional musical vocabulary, AI models need to achieve cross-modal understanding between casual lexical text and auditory perception levels. By doing so, they can better grasp the expressions conveyed by amateurs and fulfill their demands more effectively.

Previous research has primarily concentrated on specific aspects of song creation, such as converting text to music~\cite{schneider2023mo,huang2023noise2music,agostinelli2023musiclm,copet2024simple} and musical scores to songs~\cite{zhiqing2024text}, yet it has not addressed the entire creative process comprehensively. MusicGen~\cite{copet2024simple} leverages a single-stage Transformer language model to facilitate text-to-music production, employing efficient token interleaving patterns enabled by quantization-based audio codecs~\cite{zeghidour2021soundstream,defossez2022high}. Melodist~\cite{zhiqing2024text} accomplishes the generation from music score to song through a multi-stage process that includes synthesizing singing voices from inputted lyrics and melodies, creating accompaniments based on these voices, and blending the vocal tracks with the instrumental accompaniment. Concurrently, platforms specializing in description-to-song generation from the industrial sector have gained considerable attention owing to their impressive functionalities. Suno~\footnote{https://www.suno.ai}~\cite{yu2024suno} and SkyMusic~\footnote{https://www.tiangong.cn} adopt two-stage generation methods based on language and diffusion models; however, they do not grant access to their datasets and code bases. Udio~\footnote{https://www.udio.com} and Stable Audio~\footnote{https://www.stableaudio.com} opt for a single-stage diffusion model approach to undertake this task. Nevertheless, they still confront challenges such as inadequate comprehension of input text descriptions and a lack of structure in the generated songs.

To bridge this research gap, we introduce the task of \textit{Colloquial Description-to-Song Generation} aimed at the human-machine alignment in automatic song composition. This departs from traditional automatic songwriting in that: 1) It necessitates a comprehensive understanding of Colloquial Descriptions since people typically employ colloquial language instead of professional descriptions; 2) It demands a deep capture of a song's musical structure, including musical sections and rhyming structures. These unique characteristics present several challenges in colloquial description-to-song generation: 

1) Data Scarcity. They require high-quality and large-scale paired datasets that encompasses various dimensions such as colloquial descriptions, musical structure, genre, and emotion. Despite the numerous datasets proposed thus far, including those derived from automatic annotations using Music Information Retrieval (MIR) algorithms or Large Language Models (LLMs)~\cite{bertin2011million,wang2020pop909,lu2023musecoco,huang2023noise2music,melechovsky2023mustango}, as well as manual annotations~\cite{yang2017midinet,bogdanov2019mtg,wang2022songdriver,schneider2023mo,zhu2023ernie,agostinelli2023musiclm}, these existing datasets exhibit several issues that hinder their effectiveness for colloquial description-to-song generation. Firstly, current manually annotated datasets are often limited to expert annotations and have a narrow descriptive scope, which significantly diverges from the more varied descriptions provided by the general public~\cite{amer2013older,mikutta2014professional}. Secondly, there is a significant semantic gap between datasets obtained through MIR algorithms and complex human descriptions. Thirdly, due to limitations in algorithm performance, automatic annotation-based datasets cannot achieve complete accuracy, and existing manually annotated datasets, where each entry is annotated by only one person, are prone to inaccuracies caused by human errors or biases.

2) Human-Machine Alignment. It is far from enough for current models to fully understand colloquial descriptions provided by common people and to accurately learn the alignment between people's descriptions and model's behaviours. Thus, in addition to the above paired datasets, an effective framework for colloquial description-to-song generation should be proposed to capture the human-machine alignment.

    

Specifically, this study adopts supervised learning techniques alongside a dual perspective of professional and amateur annotators, and constructs the training process using an end-to-end single-stage approach. This ensures that the AI-generated song harmonizes with human desires, while also enabling the AI model to comprehend and incorporate the standard structure of human song.

To address the challenge of data scarcity, we first create a music annotation platform, named Caichong Multitask Annotation Platform (CaiMAP)~\cite{wang2024muchin}, that facilitates efficient annotation of both professional and colloquial music descriptions and employs a multi-person, multi-stage quality assurance process to guarantee the precision and uniformity of annotations.We have invited groups of professional and amateur annotators who employ two different sets of tags when labeling but annotate the same song on this platform. This leads to two distinct perspectives for each song, culminating in a comprehensive, highly precise dataset that aligns with public consensus: the Caichong Music Dataset (CaiMD)~\cite{wang2024muchin}. This dataset stands as the first open-source repository for Chinese colloquial music descriptions, covering various dimensions such as colloquial descriptions (style, emotion, etc.) and musical structure, designed to furnish exhaustive data for the fine-tuning of end-to-end models.

To handle the challenge of human-machine alignment, we propose a novel single-stage framework for colloquial description-to-song generation known as MuDiT/MuSiT. 
First, we adopt QLoRA~\cite{dettmers2024qlora} as the parameter-efficient fine-tuning (PEFT) method, and fine-tune a LLM model Qwen-14B~\cite{bai2023qwen}   to generate lyrics with additional information, including musical sections and rhyming structure, from the corresponding colloquial description. For lyrics and additional structural information, we apply a cross-attention mechanism, treating them as conditions to capture the correlations between the lyrics and the audio. 
Given that the vocabulary and phrases in the Chinese colloquial descriptions have not appeared in the texts used for training existing open-source text-to-audio comparison pre-trained models, we have trained a MuChin Cross-Modal Encoder utilizing the CaiMD dataset, modeled after architectures analogous to CLAP~\cite{wu2023large} and MuLan~\cite{huang2022mulan}. 
Subsequently, We employ MuChin to process the text description as a conditional input, standardizing the vector length before concatenating it with random noise. We then use these vectors as input, employ transformer-based diffusion models (DiT~\cite{peebles2023scalable} and SiT~\cite{ma2024sit}), that operate in the latent space of the Variational Autoencoders (VAE)~\cite{kingma2013auto}, to generate high-quality, well-structured songs that align with colloquial descriptions. For lyrics and additional structural information, we apply a cross-attention mechanism, treating them as conditions to capture the correlations between the lyrics and the audio.  And the utilization of VAE and HIFI-GAN~\cite{kong2020hifi} for decoding the song content into Mel spectrograms and converting them into WAV audio files.

In training, we initially perform supervised pre-training on DiT/SiT using a private large-scale lyrics-song audio paired dataset and unsupervised pre-training for VAE and HIFI-GAN.
Subsequently, on the task of colloquial description-to-song, the DiT/SiT are fine-tuned based on the CaiMD dataset. This process enhances their capability to generate well-structured songs that align with human colloquial descriptions.

In terms of experimental metrics, we have adopted the Fr\'echet Audio Distance (FAD) as a measure to evaluate the quality of the music produced by large-scale music models. We have also developed new metrics to assess the alignment between songs generated by music models and their colloquial descriptions, as well as to evaluate the structure of the generated lyrics. Additionally, considering the unique aspects of music, an amalgamated approach of subjective and objective evaluations was adopted.

In our main experiments, we conducted comparisons with contemporary leading large-scale music models. The results demonstrate that our MuSiT and MuDiT models exhibit superior performance in comparison with open-source frameworks such as MusicGen and Stable Audio when trained on data sets of equivalent sizes. As compared to proprietary music generation models like Suno and Udio, our models are nearly on par in terms of generation quality but present better alignment with colloquial descriptions of songs, successfully mirroring non-expert descriptors. When taking into account our discrepancy in data volume and parameter size relative to these counterparts, our models maintain their standing as the most optimal.
The performance disparity between MuSiT and MuDiT corroborates our hypothesis: SiT, with its superior capabilities in handling time-series data—particularly adept at managing continuous temporal variations and complex dynamic processes—is more suitable for music generation compared to DiT.

Additionally, supplementary experiments were carried out. We conducted an experimental evaluation of existing LLMs for the task of structured lyric generation, including Qwen~\cite{bai2023qwen}, Baichuan-2~\cite{yang2023baichuan}, GLM-130B~\cite{zeng2022glm}, and GPT-4~\cite{achiam2023gpt}, as well as a version of Qwen that was fine-tuned on the CaiMD dataset. The experiments demonstrate that the fine-tuned Qwen, despite having fewer parameters, significantly outperformed larger closed-source baseline models in structural scores, successfully aligning with the human perception of the song structure in pop music.
We also verified that MuChin, in contrast to comparative pre-trained models like MuLan and CLAP that were not fine-tuned on colloquial description-to-audio data, could effectively enhance the alignment of the output songs from MuSiT and MuDiT with colloquial descriptions.
In response to evaluating the advanced nature of the CaiMD dataset, we also designed experiments to test its efficacy. Assuming identical pre-training data, we utilized other existing text-to-music datasets to conduct fine-tuning training separately. Results revealed that models fine-tuned with the CaiMD dataset vastly surpassed others in their capacity to interpret colloquial descriptions, capturing the intentions and emotional nuances of typical users with greater precision. This validation supports a widely recognized notion, ``A small, well-annotated dataset often outperforms a larger, poorly annotated one in model training."~\cite{zhou2024lima,touvron2023llama}
In the final part of this paper, we conducted an experiment to assess whether there is a significant disparity in understanding and description of music between professionals and amateurs. The results indicate that these two groups demonstrate varied levels of interpretive differences across different dimensions and types of music, with the perspectives of professionals not resonating with amateur enthusiasts.

Our contributions are summarized as follows:
\begin{itemize}[left=1.5em]
    \item We introduce a novel task of colloquial description-to-song generation, an essential aspect of alignment works between generative AI and humans in the field of art. This task necessitates that AI models fully comprehend human colloquial descriptions to generate songs that precisely meet human expectations.
    \item We created the CaiMAP, implementing a multi-person, multi-stage quality assurance process to guarantee the precision and uniformity of annotations. Based on this platform, the CaiMD was constructed through manual annotations made from both professional musicians and amateurs, offering distinct perspectives. Unlike other existing public datasets, CaiMD is better suited for fine-tuning end-to-end colloquial description-to-song generation models, thereby aligning with amateur expressions and fulfilling the general public's demands. We are in the process of progressively open-sourcing this dataset.
    \item We propose a novel single-stage framework for colloquial description-to-song generation, referred to as MuDiT/MuSiT, to achieve human-machine alignment. The cross-modal comprehension between colloquial descriptions and auditory musical perceptions achieved by MuDiT/MuSiT enables it to align with the general public during song generation. Moreover, in addition to utilizing a fine-tuned LLM for lyrics creation, MuDiT/MuSiT exclusively employs one DiT/SiT model for end-to-end generation of other musical aspects such as melody, vocals, harmony, rhythm, and instrumentation, as well as subsequent mastering processes. This integrated approach ensures that all generated musical components are sonically harmonious and cohesive.
\end{itemize}

\section{Background}

\subsection{Text-to-Song Generation}
Text-to-song generation aims to create song pieces based to the descriptive text. Previous works typically focus on specific aspects of song creation, such as text-to-music~\cite{schneider2023mo,huang2023noise2music,agostinelli2023musiclm,copet2024simple}, lyrics-to-melody~\cite{ju2021telemelody,sheng2021songmass} and music score-to-song~\cite{zhiqing2024text}, but do not encompass the entire creative process. 
For text-to-music generation, models such as MusicLM~\cite{agostinelli2023musiclm} and MusicGen~\cite{copet2024simple} utilize quantization-based audio codecs~\cite{zeghidour2021soundstream,defossez2022high} to obtain residual codebooks and employ language models and diffusion models for high-quality audio music generation. 
For lyrics-to-melody generation, SongMASS~\cite{sheng2021songmass} and TeleMelody~\cite{ju2021telemelody}
take lyrics as input and output melodies, ensuring complex and subtle lyric-melody correlations. For music score-to-song generation, Melodist~\cite{zhiqing2024text} adopts a two-stage method that consists of singing voice synthesis from given music score and vocal-to-accompaniment synthesis based on the singing voice and natural language prompts. Meanwhile, text-to-song generation platforms from industry, such as Suno, SkyMusic, Stable Audio, and SongR~\footnote{https://www.songr.ai}, have attracted wide attention due to their impressive capabilities. Stable Audio implements an autoencoder to compress waveforms, a T5-based text embedding for text conditioning, and a transformer-based diffusion model to generate stereo songs. 
However, these platforms usually do not provide access to their datasets and methodologies, which may limit reproducibility and pose challenges for future research.

\subsection{Annotated Music Datasets}
\subsubsection{Automatic Annotation}
Automatic annotation-based datasets employ existing MIR algorithms to extract musical attributes from symbolic music or audio music. Then the extracted attributes are either incorporated into complete descriptive texts or regarded as descriptive tags. MSD~\cite{bertin2011million} collects a million of music data, along with audio, MIDI, and tags retrieved by Echo Nest Analyze API (MIR toolkit). POP909~\cite{wang2020pop909} presents a dataset containing audio, lead sheets, and other music attributes like keys and beats. MuseCoco~\cite{lu2023musecoco} and Mustango~\cite{melechovsky2023mustango} extract features from the original audio and then utilize ChatGPT to incorporate them as descriptions. MuLaMCap in Noise2Music~\cite{huang2023noise2music} utilizes an LLM to generate a set of music descriptive texts, and then employs MuLan~\cite{huang2022mulan}, a text-music embedding model to match these texts with music audio in the datasets. 
Despite this, there is a considerable semantic gap between datasets obtained through automatic annotation and complex human descriptions, which will reduce the accuracy of the datasets and limit model's performance.

\subsubsection{Manual Annotation}
Manual annotation-based datasets collect descriptions or tags from music websites, while others include data annotated by professional musicians. Hooktheory~\footnote{https://www.hooktheory.com} is a music website where users upload audio with their annotations such as melodies, chords, and beats. MTG~\cite{bogdanov2019mtg} and Mousai~\cite{schneider2023mo} use corresponding tags of music on music websites as descriptive tags, while ERNIEMusic~\cite{zhu2023ernie} uses comments of music as music descriptions, and establish datasets upon these. MusicLM~\cite{agostinelli2023musiclm} presents a dataset, MusicCaps, including music descriptions annotated by professional musicians. However, current datasets annotated manually are confined to expert annotations and limited descriptive scopes, which significantly diverge from the descriptions provided by the general public~\cite{amer2013older,mikutta2014professional}. And existing manual annotation-based datasets, where each entry is annotated by only one person, can also be prone to inaccuracies caused by human errors or biases. 

\subsection{Transformer-Based Diffusion Models}

Traditional diffusion models generally employ U-Net architectures, which are limited by the inductive biases of CNNs that struggle to effectively model the spatial correlations of signals and are not sensitive to scaling laws~\cite{li2024scalability}. However, transformer-based diffusion models (DiT)~\cite{peebles2023scalable} successfully overcome these limitations and have shown significant advantages in areas such as speech generation~\cite{liu2023vit}, image generation~\cite{bao2022all,peebles2023scalable,chen2024pixart}, and video generation~\cite{brooks2024video}. Meanwhile, scalable interpolant transformers (SiT)~\cite{ma2024sit}, built on the backbone of DiT, employ a flexible interpolant framework that connects two distributions more effectively than standard diffusion models, achieving remarkable results in efficiency and performance. However, the application of these transformer-based diffusion models to text-controlled song generation on large-scale audio datasets has yet to be verified, and their capacity to adapt to additional control information remains unresolved.
\section{Methodology}

This study adopts supervised learning techniques alongside a dual perspective of professional and amateur annotators, and constructs the training process using an end-to-end single-stage approach. This ensures that the AI-generated song harmonizes with human desires, while also enabling the AI model to comprehend and incorporate the standard structure of human songs.

\subsection{Task Definition}

In this paper, we present a novel generation task, \textit{Colloquial Description to Song}, which aims to create songs based on given colloquial descriptive text. Unlike previous works that focus on specific aspects of text-to-song generation, our newly defined task encompasses the entire creative process, from abstract description to full song production. Additionally, it possesses distinctive characteristics different from other tasks:

\begin{itemize}[left=1.5em]
    \item \textbf{Fully understanding colloquial language.} As people typically use colloquial rather than professional descriptions, it is essential to support both colloquial and professional descriptions. However, colloquial descriptions are more challenging to understand and process, necessitating the availability of relevant datasets and models for high-quality generation.
    \item \textbf{Sufficiently capturing musical structures of songs.} The musical structure of a song primarily includes musical sections and rhyming structure. And generating songs with musical structures is crucial for ensuring coherence and overall quality.
\end{itemize}

\subsection{Caichong Music Dataset Construction}

\subsubsection{CaiMAP: Caichong Multitask Annotation Platform}

We developed the CaiMAP~\cite{wang2024muchin} which adopts a novel multi-stage, multi-annotator assurance approach to enhance the accuracy of music annotations and their correspondence with commonly understood semantics. 

By engaging both amateurs and professionals in the annotation process, we ensured a robust representation of perspectives. This approach facilitated the creation of the CaiMD~\cite{wang2024muchin}, a comprehensive dataset characterized by multi-dimensional and high-precision annotations that integrate both professional and colloquial descriptions.

\subsubsubsection{1) Interface of CaiMAP.}

To bring the complex designs to life, we developed the CaiMAP, which harmonizes this series of tasks and systems. This section will provide a brief overview of the platform. 

\noindent{\textbf{Account and Login.}}  The platform features a sophisticated access control system that allocates specific roles to each user account. Users are able to log into their accounts to review and perform their assigned tasks, and subsequently submit their results for evaluation.

\noindent{\textbf{Annotation Interface.}}  After logging in and selecting a piece of music, annotators access a dedicated interface tailored for the task. This interface features a media player with adjustable progress bar and playback speed, alongside a specialized text box. It also includes a comprehensive lexicon and search tool, allowing users to select or search for descriptive terms needed for music annotation.

\noindent{\textbf{Quality Assurance Interface.}}  Upon logging in and selecting a piece of music, quality assurance inspectors are directed to a specialized interface. The interface displays annotations from two users side-by-side, highlighting differences to facilitate easy comparison. Inspectors can choose the correct annotation, make adjustments, or opt to re-annotate the piece. For other tasks, the interface shows a single annotation for the inspector to verify and score. Inspectors simply review the annotation and submit their scores.

\noindent{\textbf{Administrator Interface.}} Administrators can view submissions from any user, including annotators and quality assurance inspectors. Both the annotation and quality assurance interfaces feature a feedback button to report platform issues, allowing users to easily communicate with administrators for resolution.

\begin{figure*}[htb]
    \centering
    \includegraphics[width=\textwidth]{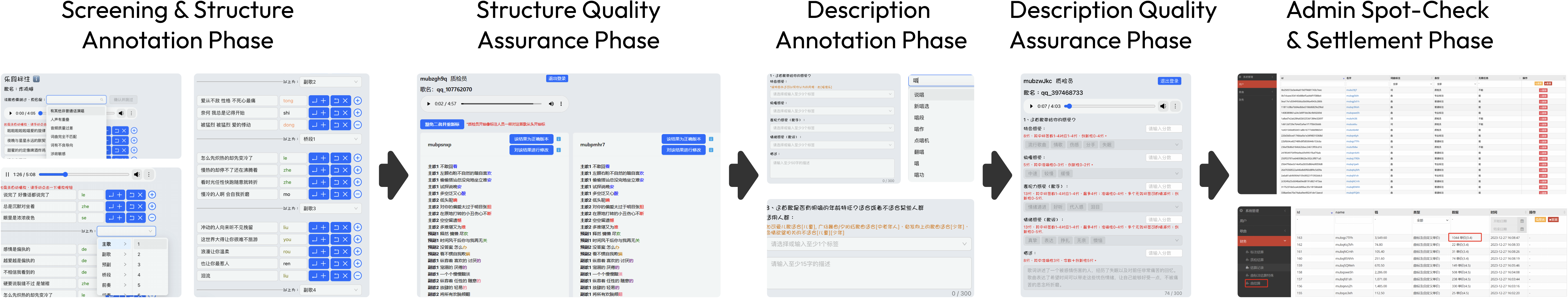}
    \caption{Pipeline of data annotation and assurance. Each annotated data undergoes 5 complex phases to ensure the accuracy. The figure shows the actual screenshots of the pages for each phase.}
    \label{fig:pipeline}
\end{figure*}

\subsubsubsection{2) Annotation and Assurance Pipeline.}

The specific annotation pipeline is shown as Fig.~\ref{fig:pipeline} and will be introduced in this section. 

\noindent{\textbf{Screening \& Structure Annotation Phase.}} In the screening phase, annotators are required to screen the data carefully. Music pieces with poor audio quality or content involving pornography or violence that are unsuitable for the dataset should be skipped.

In the structure annotation phase, the platform presents the complete lyrics sentence by sentence, and annotators are required to insert musical section tags between the lyrics. Annotators are also required to check the accuracy of the pre-annotated phonemes and rhymes for each line. If any inaccuracies are found, they should provide their own annotations.

\noindent{\textbf{Structure Quality Assurance Phase.}} To ensure the accuracy of the annotations, we implemented a quality assurance mechanism. Each piece of data undergoes annotation by two separate annotators. Subsequently, the platform autonomously verifies the congruence of the annotations. If they align, the platform seamlessly integrates the data into the dataset for the subsequent phase. In instances of disparities, both sets of annotations are referred to a quality assurance inspector for resolution. The inspector determines the correct annotation or submits an independent correction if necessary.

\noindent{\textbf{Description Annotation Phase.}} Data that successfully clears the structure quality assurance phase becomes eligible for utilization in the music description phase. During this phase, to guarantee attentive listening and thoughtful music descriptions, annotators must listen to each music piece without interruption. Specifically, annotators are prohibited from writing any textual descriptions within the initial 30 seconds of the music piece. Copy and paste content is also not allowed. Additionally, limitations are imposed on the number of tags that can be entered and on the word count of user-generated entries.

\noindent{\textbf{Description Quality Assurance Phase.}} Since music description annotation involves subjective judgments and is challenging to assess, the platform employs a randomized selection process, choosing 20\% of the annotation results from each annotator for submission to quality assurance inspectors for scoring. These scores are then logged in the platform’s backend. Annotated data that successfully pass the sampling quality assurance are submitted into the dataset, whereas those that do not meet the standards are rejected.

\noindent{\textbf{Admin Spot-Check \& Settlement Phase.}} Administrators can monitor the real-time progress of each group’s work and make payments accordingly, depending on the outcomes of quality assurance checks. Annotators who consistently achieve high pass rates for their annotations will be rewarded additionally, whereas those with lower pass rates will incur penalties, thus motivating them to annotate diligently.

To determine whether the inspectors are competent in their work, administrators also have the access to randomly selected samples of their work for secondary verification.

\subsubsection{Annotation Process}
To effectively annotate music with both amateur and professional descriptions, we have engaged 213 individuals familiar with Chinese music, comprising 109 amateur enthusiasts and 104 professionals. This diverse group was recruited through campus and public efforts and includes 144 males and 69 females, aged 19 to 35 years. We have organized these participants into four groups, each assigned specific tasks as follows: 

\begin{itemize}[left=1.5em]
    \item \textbf{Professional Group.} Annotate musical sections, rhyming structure and provide professional descriptions.
    \item \textbf{Amateur Group.} Provide colloquial descriptions.
    \item \textbf{Inspector Group.} Evaluate structure annotations, and score music descriptions.
    \item \textbf{Administrator.} Address and provide feedback on inquiries from various groups, and conduct random spot-checks of the groups’ outcomes.
\end{itemize}




\begin{table}[htbp]
\centering
\caption{Classification of Annotation Tasks.}
\label{tab:class_task}
\begin{tabularx}{\textwidth}{X X}
\hline
\textbf{Classification} & \textbf{Task}                  \\ \hline
\multirow{4}{*}{A}      & Musical Sections Annotation     \\
                        & Lyrics Correction               \\
                        & Lyrics Screening                \\
                        & Rhyming Structure Annotation               \\ \hline
\multirow{2}{*}{B}      & Professional Music Description \\
                        & Amateur Music Description      \\ \hline
\end{tabularx}
\end{table}

\subsubsubsection{1) Quality Assurance Mechanisms.}
\noindent{\textbf{Annotation Task Classification.}} Annotation tasks are categorized into two types: Type A (objectively assessed) and Type B (subjectively assessed) as shown in Table~\ref{tab:class_task}. For Type A tasks, two annotators are assigned per song, while Type B tasks require only one. The tasks are completed in phases: initially, the Structure Annotation Phase (Type A), followed by the Music Description Annotation Phase (Type B). Only annotations that pass quality assurance in each phase are included in the dataset. Discrepancies in Type A are resolved by inspectors who determine the accuracy or correct errors, whereas in Type B, inspectors score annotations on a scale of 0-100.

\noindent{\textbf{Classification of Annotators.}} Annotators are rigorously screened and grouped based on accuracy or scores. Those with persistent low performance receive warnings or are excluded from future tasks. High performers are rewarded, and intermediate performers may face penalties.

\noindent{\textbf{Additional Quality Assurance Measures.}} For content suitability, annotators exclude data with inappropriate content or poor quality. During Type A tasks, engagement is measured by time spent and interactions with the music player. For Type B, annotators must listen attentively to the entire song, avoiding premature interactions with the platform or copying text.

\subsubsubsection{2) Individual Grouping and Training.} 
Detail the grouping and training method for each group of individuals.

\noindent{\textbf{Grouping.}} During the structure annotation phase, involving Type A tasks, each data piece requires dual annotations, engaging 104 professionals. From these, 11 individuals are selected as quality assurance inspectors based on their exceptional expertise and detailed evaluations of their resumes. The remaining 93 serve as annotators. In the music description annotation phase, Type B tasks require only one annotation per data piece, thus involving fewer participants. Here, 109 amateurs annotate using colloquial terms, while the same 93 professionals provide more technical descriptions. The 11 inspectors continue their roles, and an experienced, communicative member from our team is appointed as the platform administrator to oversee the process.

\noindent{\textbf{Training.}} Annotators start by pre-annotating a small dataset of 20 entries, learning platform features and proper annotation practices, including correcting common errors. Inspectors receive more in-depth training, mastering the platform and developing consistent evaluation standards. They review the same dataset after annotators, making judgments based on guidelines and adjusting as needed. Discrepancies in scores over 10 points are discussed in meetings to standardize evaluations. This cycle continues until inspectors consistently align in their assessments.

\subsubsubsection{3) Automatic Annotation.}
The performance of algorithms designed for annotating textual descriptions, lyrics, and musical sections is often unsatisfactory due to their dependence on subjective human evaluations. Additionally, other types of data such as phonetic alignment, vocal separation, and audio-to-MIDI conversion do not reliably match human perception. Manual annotation of these elements is challenging and time-consuming. 

However, advanced algorithms now exist that efficiently handle these tasks, as detailed in next section. Consequently, we employ data pre-processing algorithms for automatic annotation, eliminating the need for manual effort and seamlessly integrating this processed content into our dataset.

\noindent{\textbf{Music Genre Clustering.}} To reduce subjective bias and promote diverse descriptions across music genres, we employ MERT~\cite{li2023mert}, a pre-trained music audio encoder, to process and cluster the encoded data into 1000 unique audio clusters. This data is then evenly distributed among annotators, ensuring a balanced variety of music genres for labeling. This method ensures that each cluster is annotated by a diverse range of annotators, greatly enhancing the diversity and depth of the annotated data.

\noindent{\textbf{Vocal \& Track Separation.}} To prepare the dataset for tasks like accompaniment generation, melody generation, and vocal synthesis, we use Demucs~\cite{rouard2022hybrid,defossez2021hybrid} to separate vocals from the musical accompaniment in audio files. Additionally, we isolate individual instrument tracks, including drums and bass, to meet the needs of a broader range of music-related tasks.

\noindent{\textbf{Phonemic Level Alignment in Audio-Lyrics.}} To prepare audio-lyrics pairs for vocal synthesis, we use the Montreal Forced Aligner (MFA)~\cite{mfa} to achieve phonemic level alignment.

\noindent{\textbf{Automatic Pre-Annotation.}} To enhance the efficiency of future manual annotations, we've introduced software for automatic pre-annotation of lyric-related tasks. This includes a specialized program for pre-annotating rhyming structure and a fine-tuned version of Qwen for identifying the main themes in lyrics. These pre-annotations provide a foundation for manual review, allowing annotators to assess and refine the automatic annotations or use them as guidelines for their annotation efforts.

\noindent{\textbf{Lead Sheet Transcription.}} To support symbolic music tasks using MIDI, we transcribe audio into lead sheets using Sheet Sage~\cite{donahue2021sheet}, software that employs the Jukebox encoding model~\cite{dhariwal2020jukebox}. 

\begin{figure*}[htb]
    \centering
    \includegraphics[width=1.05\textwidth]{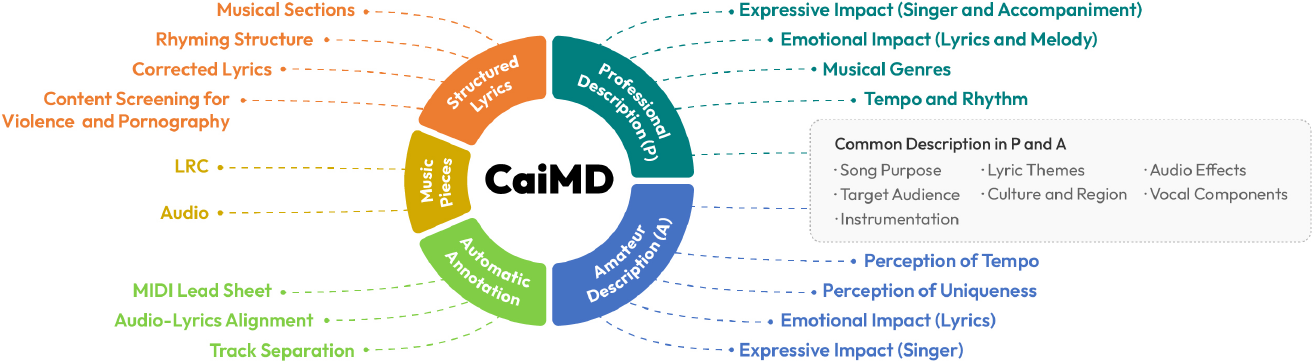}
    \caption{An overview of CaiMD. The Chinese Colloquial Descriptions consist of Description(A) and Common Description(P and A) annotated by amateur annotators. In addition, we recruit professional annotators to label Description(P), Musical Sections, and Rhyming Structures of the lyrics. And machine-annotated information such as MIDI is also incorporated.}
    
    \label{fig:CaiMD}
\end{figure*}

After the aforementioned processes, the corpus encompassed by CaiMD is depicted in Fig.~\ref{fig:CaiMD}.

\subsection{MuDiT/MuSiT Framework}

\begin{figure*}[h]
    \centering
    \includegraphics[width=0.7\textwidth]{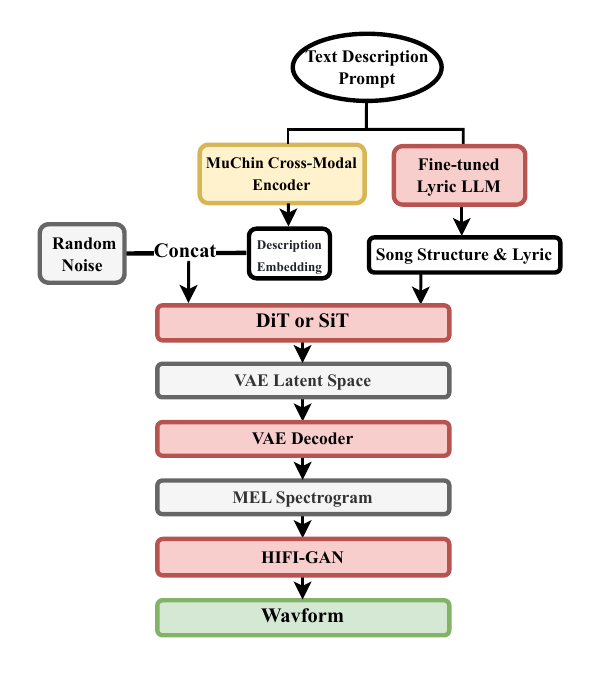}
    \caption{The overall architecture of MuDiT/MuSiT. We transform the text description into vectors using MuChin, which are then concatenated with noise to serve as the noised latent input for DiT. Additionally, we use a fine-tuned LLM to generate structured lyrics. Next, DiT/SiT generates the entire song content in the form of VAE latent space, encompassing melody, harmony, rhythm, vocals, instrumentation, and other musical components. Finally, we use VAE decoder and HIFI-GAN to decode the song content into Mel spectrograms and convert them into WAV audio files.}
    \label{fig:system_overview}
\end{figure*}

\subsubsection{Framework Overview}

We propose a novel end-to-end generation model called MuDiT/MuSiT that converts colloquial descriptions into songs. The MuDiT/MuSiT comprises several components: a fine-tuned LLM designed to generate structured lyrics; MuChin for cross-modal text-to-audio encoding; transformer-based diffusion models (DiT~\cite{peebles2023scalable} and SiT~\cite{ma2024sit}) to generate songs that align with colloquial descriptions; and the use of VAE and HIFI-GAN~\cite{kong2020hifi} to decode the song content into Mel spectrograms and convert them into WAV audio files.

First, the system uses a fine-tuned LLM to generate structured lyrics from the user's colloquial descriptions. The generated lyrics serve as conditional input to the cross-attention module of the DiT model. Next, the system employs MuChin for cross-modal text-to-audio encoding, converting the user's text descriptions into description embeddings. These embeddings, concatenated with random noise, are fed into the DiT model and undergo cross-attention learning with the lyrics. Following this, the system integrates transformer-based diffusion models, namely DiT and SiT, which are responsible for generating VAE latent variables. These VAE latent variables are used as indices for audio sampling in the VAE audio space. Finally, the system uses VAE and HIFI-GAN to decode the generated song content into Mel spectrograms and convert these spectrograms into high-quality WAV files. The VAE and HIFI-GAN work together to maintain the fidelity and richness of the audio, ensuring that the final output is a polished and professionally sounding song.

\subsubsection{Transforming Noise to VAE Space}

\begin{figure*}[h]
    \vspace{-2em}
    \centering
    \includegraphics[width=0.7\textwidth]{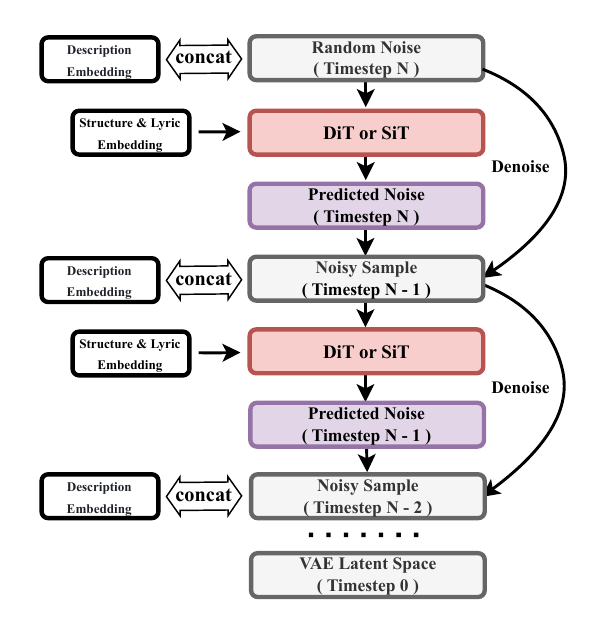}
    \caption{The complete diffusion process of DiT/SiT. The DiT/SiT model starts at timestep t = N and repeatedly subtracts the predicted noise from the noisy sample, continuing this process multiple times until it reaches timestep t = 0, resulting in the final song content, which is output in the form of a VAE latent space.}
    \label{fig:noise2vae}
\end{figure*}

In the original audio data space, each point represents meaningful musical content. We aim for an audio space that resembles a simple distribution, such as a normal distribution. However, the original audio data space is complex and high-dimensional, necessitating an audio space transformation to convert the irregular space into a regular distribution.

During training, we first perform unsupervised pre-training on the VAE and HIFI-GAN using a large private dataset. By using the VAE encoder, we transform the song audio into VAE latent vectors. The VAE audio data space acts like a dictionary with a distribution similar to a normal distribution, and the VAE latent vectors serve as indices to locate the corresponding audio content within the VAE space. Therefore, the training data for DiT needs to be a subset of the VAE training data.

Since the VAE latent space is inherently discontinuous and discrete, not every position contains meaningful musical content. Noise is distributed in the meaningless positions between the discrete points. During inference of the DiT/SiT, the original random noise does not lie within the VAE latent space. However, by iteratively subtracting the noise, it gradually approaches a meaningful latent point within the VAE latent space. Subsequently, we use the VAE decoder to convert this latent point into a MEL spectrogram, and then use HIFI-GAN to transform it into a waveform.

\subsubsection{MuChin Cross-Modal Encoder}

The text-audio contrastive pre-training model is crucial for the AI's understanding of colloquial descriptions and cannot be replaced with other text encoders trained on professional vocabulary. 

Given that the vocabulary and phrases in the Chinese colloquial descriptions have not appeared in the texts used for training existing open-source text-to-audio comparison pre-trained models, we have trained the MuChin utilizing the CaiMD dataset, modeled after architectures analogous to CLAP~\cite{wu2023large} and MuLan~\cite{huang2022mulan}. MuChin is a cross-modal encoder for word-audio pairs. When the user inputs text description prompts, MuChin converts them into description embddings, which are then concat with random noise and fed into the DiT model. This step ensures that the input text description prompt is transformed into a dense vector representation that captures the semantic nuances necessary for music generation.

During the training of the MuChin model, we randomly segmented the lengths of professional and colloquial descriptions in the CaiMD dataset, as well as the lengths of the song audio. This allows MuChin to perform text-audio contrastive pre-training effectively, adapting to inputs of varying text and audio lengths.

\subsubsection{Fine-tuned Lyric LLM}

Based on considerations of model parameters, we selected the provenly effective Qwen-14B-Chat-Int4~\cite{bai2023qwen} model for lyric generation within the context of Mandarin Chinese. Our training data encompass themes extracted from lyrics along with manually annotated verses featuring musical sections and rhyming structures. This dataset was employed to fine-tune the model for the task of generating lyrics that included musical sections and rhyming structure from text prompts.

Specifically, we converted Chinese characters into pinyin to serve as the text input for the DiT model, enhancing its generalization capabilities and robustness. The generated outcomes comprise tags for musical sections such as \textless{}verse\textgreater{}, \textless{}chorus\textgreater{}, and \textless{}bridge\textgreater{}. The DiT model is responsible for creating songs that adhere to these labeled musical structures.

\subsubsection{Control Conditions for DiT/SiT}

We applied DiT to the task of song generation, adopting this new standardized architecture to open up more possibilities for cross-domain research. 

\begin{figure*}[h]
    \centering
    \includegraphics[width=0.9\textwidth]{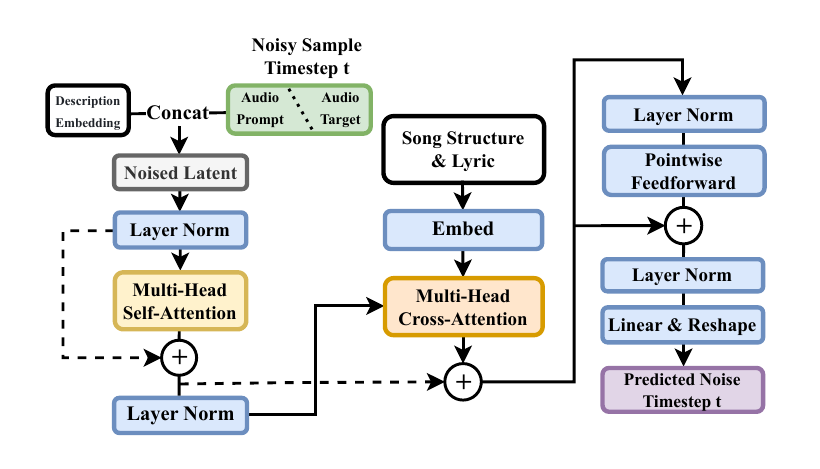}
    \caption{One timestep in the diffusion process of DiT/SiT. The noisy sample is Random Noise only at timestep t = N; in subsequent timesteps, it is the denoised sample. Due to the varying length characteristic of song structures and lyrics, they can only be processed through cross-attention. In contrast, text descriptions are converted into fixed-length vectors via MuChin, allowing them to be concatenated with the noisy sample and processed through self-attention. Finally, the predicted noise at the current timestep t is output.}
    \label{fig:control_conditions}
\end{figure*}

\subsubsubsection{1) Application of Self-Attention Mechanism.}

\noindent{\textbf{Text Description Prompt Condition.}} We utilize MuChin to process text descriptions into description embeddings. After normalizing the vector lengths, we concatenate these embeddings with noisy samples. The concatenated vectors serve as noised latents, which are then processed by the multi-head self-attention mechanism in DiT/SiT. This multi-head self-attention mechanism enables the model to concurrently attend to different segments of the input latent vectors, effectively capturing the complex dependencies between text descriptions and song audio content.

\noindent{\textbf{Audio Prompt Condition.}} The noisy sample in Fig.~\ref{fig:control_conditions} is divided into two parts: the prompt and the target. The prompt part boasts high extensibility. It can accept user-supplied a cappella, instrumental solos, or reference song audios to serve as controlling conditions, enabling the continuation and generation of songs containing the corresponding content or style. Interestingly, sounds like tapping on a desk or animal barks can also be input and incorporated into certain musical elements of the final song. We use source separation technology on the original song to obtain vocals, drums, chord instruments, and bass instruments as part of the training data, thereby achieving this control effect.

\subsubsubsection{2) Application of Cross-Attention Mechanism.}

\noindent{\textbf{Lyric \& Song Structure Condition.}} Due to the variable length of vectors for lyrics and additional structural information, direct processing via self-attention is not feasible. However, the presence of the cross-attention mechanism allows DiT/SiT to leverage the advantages of Transformers while retaining the functional benefits of diffusion models. This enables DiT/SiT to perform multi-head cross-attention learning with variable-length lyric content and audio content, treating them as parallel streams to capture the correlations between lyrics and audio. Specifically, we map the music section labels as a single token into the lyric dictionary rather than embedding it separately. This approach prevents the disruption of the temporal relationship between lyrics and musical sections.

\subsubsection{Training and Inference of DiT/SiT}

During the training of DiT/SiT, we employ DDPM with random timesteps, while during inference, we use DDIM with sequential timesteps (progressively from t to 0). 

\subsubsubsection{1) Pre-training Phase.}

We performed supervised pre-training on DiT/SiT using a large private dataset of paired ``lyrics-song audio''. The lyrics text serves as a supervision signal through cross-attention, while the song audio is used as training data in the form of VAE latent vectors.

Considering that during the training phase, lyrics timestamps can be used to align audio windows with corresponding lyrics. However, during the inference phase, lyrics provided by users or LLMs often lack precise timestamps. This results in the model being unable to assign appropriate lengths of lyrics text to each audio window during inference. The variability in singing speed is more pronounced compared to speech synthesis, which has relatively consistent speed, making it challenging to predict the approximate time range based on the number of words in the text.

Therefore, we ultimately decided not to adopt the window-based audio generation method commonly used in large speech generation models. Instead, we opted to generate the entire length of the song in a single pass. During the training phase, we do not randomly segment the audio; rather, we use the entire song audio as training data. To address the issue of variable length, we add padding to the end of each audio segment to standardize the length for training purposes.

This approach not only resolves the challenge of dividing lyrics text into windows but also brings two additional benefits: 1) Compared to window-based generation, it better enables the DiT/SiT model to capture the overall musical structure of the song. 2) During the training phase, there is no need for timestamped lyrics data (.LRC); only the regular lyrics data (.TXT) is required.

\subsubsubsection{2) Fine-tuning Phase.}

Finally, for the task of colloquial description to song generation, we fine-tuned the DiT/SiT model based on the CaiMD dataset. This fine-tuning enables the DiT/SiT model to generate well-structured songs that align with human colloquial expressions based on user input of colloquial descriptions and lyrics generated by the fine-tuned lyric LLM.

Considering that users input description prompts of varying lengths, from single words to full paragraphs, during inference, we ensure consistency between training and inference. To achieve this, we randomly segment the professional and amateur description annotations in the CaiMD dataset into varying lengths during the training of DiT/SiT. Each segmented word, phrase, or sentence is converted into a corresponding vector via MuChin and then concatenated along the length dimension, keeping the hidden dimension unchanged. As a result, the DiT/SiT model is exposed to description prompts of different lengths during training, allowing it to accommodate varying lengths of text inputs during inference.

\subsubsection{Differential Benefits of SiT over DiT}

The DiT sub-model can be replaced by the SiT sub-model, and both share a similar overall neural network architecture. However, SiT introduces a new interpolation framework and improves the sampling mechanism, providing a more detailed exploration of the diffusion process.

Interpolation refers to the data transformation path during the perturbation of raw data into Gaussian noise. DiT operates on discrete time steps, achieving distribution transformation by defining step-dependent discrete time decay and fixed noise coefficients, assuming adjacent distributions convert at a constant rate, making the transformation rigid. SiT generalizes this to continuous time, allowing the model to establish more flexible and smooth connections between the original data and Gaussian distributions, discarding prior assumptions in the discrete process, and choosing better-performing interpolation functions. Clearly, the continuous diffusion process aligns better with the token continuity in music. Based on this, we designed new continuous interpolation functions. Specifically, for the original data \(x^* \sim p(x)\)and noise \(\epsilon \sim N(0, 1)\), the transformation at any time \(x_t\) can be expressed as:

\[x_t = \alpha_tx^*+\sigma_t\epsilon,\]

where the coefficients are:

\[
\alpha_t = \cos\left(\frac{\pi t}{2}\right)
\]
\[
\sigma_t = \sin\left(\frac{\pi t}{2}\right)
\]

This design ensures adaptability in music generation and mitigates infringement issues.

Regarding the sampling mechanism, traditional DiT relies on a deterministic sampling process, whereas SiT introduces randomness, decoupling the diffusion coefficients between inference and training. This design reduces the risk of overfitting and improves the model's generalization capability. Specifically, SiT defines the relationship between the velocity field (rate of distribution change) and the score function (quality of generation) based on the reverse-time stochastic differential equation. This allows the estimation of complex score functions through simpler velocity fields, which can be learned by neural networks. Consequently, we can introduce significant flexibility in the music generation process, adapting to complex data distributions and avoiding the generation of mechanical-sounding music.


\bibliography{sn-bibliography}


\begin{thebibliography}{49}
\ifx \bisbn   \undefined \def \bisbn  #1{ISBN #1}\fi
\ifx \binits  \undefined \def \binits#1{#1}\fi
\ifx \bauthor  \undefined \def \bauthor#1{#1}\fi
\ifx \batitle  \undefined \def \batitle#1{#1}\fi
\ifx \bjtitle  \undefined \def \bjtitle#1{#1}\fi
\ifx \bvolume  \undefined \def \bvolume#1{\textbf{#1}}\fi
\ifx \byear  \undefined \def \byear#1{#1}\fi
\ifx \bissue  \undefined \def \bissue#1{#1}\fi
\ifx \bfpage  \undefined \def \bfpage#1{#1}\fi
\ifx \blpage  \undefined \def \blpage #1{#1}\fi
\ifx \burl  \undefined \def \burl#1{\textsf{#1}}\fi
\ifx \doiurl  \undefined \def \doiurl#1{\url{https://doi.org/#1}}\fi
\ifx \betal  \undefined \def \betal{\textit{et al.}}\fi
\ifx \binstitute  \undefined \def \binstitute#1{#1}\fi
\ifx \binstitutionaled  \undefined \def \binstitutionaled#1{#1}\fi
\ifx \bctitle  \undefined \def \bctitle#1{#1}\fi
\ifx \beditor  \undefined \def \beditor#1{#1}\fi
\ifx \bpublisher  \undefined \def \bpublisher#1{#1}\fi
\ifx \bbtitle  \undefined \def \bbtitle#1{#1}\fi
\ifx \bedition  \undefined \def \bedition#1{#1}\fi
\ifx \bseriesno  \undefined \def \bseriesno#1{#1}\fi
\ifx \blocation  \undefined \def \blocation#1{#1}\fi
\ifx \bsertitle  \undefined \def \bsertitle#1{#1}\fi
\ifx \bsnm \undefined \def \bsnm#1{#1}\fi
\ifx \bsuffix \undefined \def \bsuffix#1{#1}\fi
\ifx \bparticle \undefined \def \bparticle#1{#1}\fi
\ifx \barticle \undefined \def \barticle#1{#1}\fi
\bibcommenthead
\ifx \bconfdate \undefined \def \bconfdate #1{#1}\fi
\ifx \botherref \undefined \def \botherref #1{#1}\fi
\ifx \url \undefined \def \url#1{\textsf{#1}}\fi
\ifx \bchapter \undefined \def \bchapter#1{#1}\fi
\ifx \bbook \undefined \def \bbook#1{#1}\fi
\ifx \bcomment \undefined \def \bcomment#1{#1}\fi
\ifx \oauthor \undefined \def \oauthor#1{#1}\fi
\ifx \citeauthoryear \undefined \def \citeauthoryear#1{#1}\fi
\ifx \endbibitem  \undefined \def \endbibitem {}\fi
\ifx \bconflocation  \undefined \def \bconflocation#1{#1}\fi
\ifx \arxivurl  \undefined \def \arxivurl#1{\textsf{#1}}\fi
\csname PreBibitemsHook\endcsname

\bibitem[\protect\citeauthoryear{Bertin-Mahieux et~al.}{2011}]{bertin2011million}
\begin{botherref}
\oauthor{\bsnm{Bertin-Mahieux}, \binits{T.}},
\oauthor{\bsnm{Ellis}, \binits{D.P.}},
\oauthor{\bsnm{Whitman}, \binits{B.}},
\oauthor{\bsnm{Lamere}, \binits{P.}}:
The million song dataset
(2011)
\end{botherref}
\endbibitem

\bibitem[\protect\citeauthoryear{Yang et~al.}{2017}]{yang2017midinet}
\begin{botherref}
\oauthor{\bsnm{Yang}, \binits{L.-C.}},
\oauthor{\bsnm{Chou}, \binits{S.-Y.}},
\oauthor{\bsnm{Yang}, \binits{Y.-H.}}:
Midinet: A convolutional generative adversarial network for symbolic-domain music generation.
arXiv preprint arXiv:1703.10847
(2017)
\end{botherref}
\endbibitem

\bibitem[\protect\citeauthoryear{Bogdanov et~al.}{2019}]{bogdanov2019mtg}
\begin{bchapter}
\bauthor{\bsnm{Bogdanov}, \binits{D.}},
\bauthor{\bsnm{Won}, \binits{M.}},
\bauthor{\bsnm{Tovstogan}, \binits{P.}},
\bauthor{\bsnm{Porter}, \binits{A.}},
\bauthor{\bsnm{Serra}, \binits{X.}}:
\bctitle{The mtg-jamendo dataset for automatic music tagging}.
(\byear{2019}).
\bcomment{ICML}
\end{bchapter}
\endbibitem

\bibitem[\protect\citeauthoryear{Wang et~al.}{2020}]{wang2020pop909}
\begin{botherref}
\oauthor{\bsnm{Wang}, \binits{Z.}},
\oauthor{\bsnm{Chen}, \binits{K.}},
\oauthor{\bsnm{Jiang}, \binits{J.}},
\oauthor{\bsnm{Zhang}, \binits{Y.}},
\oauthor{\bsnm{Xu}, \binits{M.}},
\oauthor{\bsnm{Dai}, \binits{S.}},
\oauthor{\bsnm{Gu}, \binits{X.}},
\oauthor{\bsnm{Xia}, \binits{G.}}:
Pop909: A pop-song dataset for music arrangement generation.
arXiv preprint arXiv:2008.07142
(2020)
\end{botherref}
\endbibitem

\bibitem[\protect\citeauthoryear{Wang et~al.}{2022}]{wang2022songdriver}
\begin{bchapter}
\bauthor{\bsnm{Wang}, \binits{Z.}},
\bauthor{\bsnm{Zhang}, \binits{K.}},
\bauthor{\bsnm{Wang}, \binits{Y.}},
\bauthor{\bsnm{Zhang}, \binits{C.}},
\bauthor{\bsnm{Liang}, \binits{Q.}},
\bauthor{\bsnm{Yu}, \binits{P.}},
\bauthor{\bsnm{Feng}, \binits{Y.}},
\bauthor{\bsnm{Liu}, \binits{W.}},
\bauthor{\bsnm{Wang}, \binits{Y.}},
\bauthor{\bsnm{Bao}, \binits{Y.}}, \betal:
\bctitle{Songdriver: Real-time music accompaniment generation without logical latency nor exposure bias}.
In: \bbtitle{Proceedings of the 30th ACM International Conference on Multimedia},
pp. \bfpage{1057}--\blpage{1067}
(\byear{2022})
\end{bchapter}
\endbibitem

\bibitem[\protect\citeauthoryear{Huang et~al.}{2023}]{huang2023noise2music}
\begin{botherref}
\oauthor{\bsnm{Huang}, \binits{Q.}},
\oauthor{\bsnm{Park}, \binits{D.S.}},
\oauthor{\bsnm{Wang}, \binits{T.}},
\oauthor{\bsnm{Denk}, \binits{T.I.}},
\oauthor{\bsnm{Ly}, \binits{A.}},
\oauthor{\bsnm{Chen}, \binits{N.}},
\oauthor{\bsnm{Zhang}, \binits{Z.}},
\oauthor{\bsnm{Zhang}, \binits{Z.}},
\oauthor{\bsnm{Yu}, \binits{J.}},
\oauthor{\bsnm{Frank}, \binits{C.}}, et al.:
Noise2music: Text-conditioned music generation with diffusion models.
arXiv preprint arXiv:2302.03917
(2023)
\end{botherref}
\endbibitem

\bibitem[\protect\citeauthoryear{Melechovsky et~al.}{2023}]{melechovsky2023mustango}
\begin{botherref}
\oauthor{\bsnm{Melechovsky}, \binits{J.}},
\oauthor{\bsnm{Guo}, \binits{Z.}},
\oauthor{\bsnm{Ghosal}, \binits{D.}},
\oauthor{\bsnm{Majumder}, \binits{N.}},
\oauthor{\bsnm{Herremans}, \binits{D.}},
\oauthor{\bsnm{Poria}, \binits{S.}}:
Mustango: Toward controllable text-to-music generation.
arXiv preprint arXiv:2311.08355
(2023)
\end{botherref}
\endbibitem

\bibitem[\protect\citeauthoryear{Lu et~al.}{2023}]{lu2023musecoco}
\begin{botherref}
\oauthor{\bsnm{Lu}, \binits{P.}},
\oauthor{\bsnm{Xu}, \binits{X.}},
\oauthor{\bsnm{Kang}, \binits{C.}},
\oauthor{\bsnm{Yu}, \binits{B.}},
\oauthor{\bsnm{Xing}, \binits{C.}},
\oauthor{\bsnm{Tan}, \binits{X.}},
\oauthor{\bsnm{Bian}, \binits{J.}}:
Musecoco: Generating symbolic music from text.
arXiv preprint arXiv:2306.00110
(2023)
\end{botherref}
\endbibitem

\bibitem[\protect\citeauthoryear{Schneider et~al.}{2023}]{schneider2023mo}
\begin{botherref}
\oauthor{\bsnm{Schneider}, \binits{F.}},
\oauthor{\bsnm{Kamal}, \binits{O.}},
\oauthor{\bsnm{Jin}, \binits{Z.}},
\oauthor{\bsnm{Sch{\"o}lkopf}, \binits{B.}}:
Mo$\backslash$\^{} usai: Text-to-music generation with long-context latent diffusion.
arXiv preprint arXiv:2301.11757
(2023)
\end{botherref}
\endbibitem

\bibitem[\protect\citeauthoryear{Zhu et~al.}{2023}]{zhu2023ernie}
\begin{botherref}
\oauthor{\bsnm{Zhu}, \binits{P.}},
\oauthor{\bsnm{Pang}, \binits{C.}},
\oauthor{\bsnm{Chai}, \binits{Y.}},
\oauthor{\bsnm{Li}, \binits{L.}},
\oauthor{\bsnm{Wang}, \binits{S.}},
\oauthor{\bsnm{Sun}, \binits{Y.}},
\oauthor{\bsnm{Tian}, \binits{H.}},
\oauthor{\bsnm{Wu}, \binits{H.}}:
Ernie-music: Text-to-waveform music generation with diffusion models.
arXiv preprint arXiv:2302.04456
(2023)
\end{botherref}
\endbibitem

\bibitem[\protect\citeauthoryear{Sheng et~al.}{2021}]{sheng2021songmass}
\begin{bchapter}
\bauthor{\bsnm{Sheng}, \binits{Z.}},
\bauthor{\bsnm{Song}, \binits{K.}},
\bauthor{\bsnm{Tan}, \binits{X.}},
\bauthor{\bsnm{Ren}, \binits{Y.}},
\bauthor{\bsnm{Ye}, \binits{W.}},
\bauthor{\bsnm{Zhang}, \binits{S.}},
\bauthor{\bsnm{Qin}, \binits{T.}}:
\bctitle{Songmass: Automatic song writing with pre-training and alignment constraint}.
In: \bbtitle{Proceedings of the AAAI Conference on Artificial Intelligence},
vol. \bseriesno{35},
pp. \bfpage{13798}--\blpage{13805}
(\byear{2021})
\end{bchapter}
\endbibitem

\bibitem[\protect\citeauthoryear{Zeng et~al.}{2021}]{zeng2021musicbert}
\begin{botherref}
\oauthor{\bsnm{Zeng}, \binits{M.}},
\oauthor{\bsnm{Tan}, \binits{X.}},
\oauthor{\bsnm{Wang}, \binits{R.}},
\oauthor{\bsnm{Ju}, \binits{Z.}},
\oauthor{\bsnm{Qin}, \binits{T.}},
\oauthor{\bsnm{Liu}, \binits{T.-Y.}}:
Musicbert: Symbolic music understanding with large-scale pre-training.
arXiv preprint arXiv:2106.05630
(2021)
\end{botherref}
\endbibitem

\bibitem[\protect\citeauthoryear{Hsiao et~al.}{2021}]{hsiao2021compound}
\begin{bchapter}
\bauthor{\bsnm{Hsiao}, \binits{W.-Y.}},
\bauthor{\bsnm{Liu}, \binits{J.-Y.}},
\bauthor{\bsnm{Yeh}, \binits{Y.-C.}},
\bauthor{\bsnm{Yang}, \binits{Y.-H.}}:
\bctitle{Compound word transformer: Learning to compose full-song music over dynamic directed hypergraphs}.
In: \bbtitle{Proceedings of the AAAI Conference on Artificial Intelligence},
vol. \bseriesno{35},
pp. \bfpage{178}--\blpage{186}
(\byear{2021})
\end{bchapter}
\endbibitem

\bibitem[\protect\citeauthoryear{Yu et~al.}{2022}]{yu2022museformer}
\begin{barticle}
\bauthor{\bsnm{Yu}, \binits{B.}},
\bauthor{\bsnm{Lu}, \binits{P.}},
\bauthor{\bsnm{Wang}, \binits{R.}},
\bauthor{\bsnm{Hu}, \binits{W.}},
\bauthor{\bsnm{Tan}, \binits{X.}},
\bauthor{\bsnm{Ye}, \binits{W.}},
\bauthor{\bsnm{Zhang}, \binits{S.}},
\bauthor{\bsnm{Qin}, \binits{T.}},
\bauthor{\bsnm{Liu}, \binits{T.-Y.}}:
\batitle{Museformer: Transformer with fine-and coarse-grained attention for music generation}.
\bjtitle{Advances in Neural Information Processing Systems}
\bvolume{35},
\bfpage{1376}--\blpage{1388}
(\byear{2022})
\end{barticle}
\endbibitem

\bibitem[\protect\citeauthoryear{Agostinelli et~al.}{2023}]{agostinelli2023musiclm}
\begin{botherref}
\oauthor{\bsnm{Agostinelli}, \binits{A.}},
\oauthor{\bsnm{Denk}, \binits{T.I.}},
\oauthor{\bsnm{Borsos}, \binits{Z.}},
\oauthor{\bsnm{Engel}, \binits{J.}},
\oauthor{\bsnm{Verzetti}, \binits{M.}},
\oauthor{\bsnm{Caillon}, \binits{A.}},
\oauthor{\bsnm{Huang}, \binits{Q.}},
\oauthor{\bsnm{Jansen}, \binits{A.}},
\oauthor{\bsnm{Roberts}, \binits{A.}},
\oauthor{\bsnm{Tagliasacchi}, \binits{M.}}, et al.:
Musiclm: Generating music from text.
arXiv preprint arXiv:2301.11325
(2023)
\end{botherref}
\endbibitem

\bibitem[\protect\citeauthoryear{Wu et~al.}{2023}]{wu2023melodyglm}
\begin{botherref}
\oauthor{\bsnm{Wu}, \binits{X.}},
\oauthor{\bsnm{Huang}, \binits{Z.}},
\oauthor{\bsnm{Zhang}, \binits{K.}},
\oauthor{\bsnm{Yu}, \binits{J.}},
\oauthor{\bsnm{Tan}, \binits{X.}},
\oauthor{\bsnm{Zhang}, \binits{T.}},
\oauthor{\bsnm{Wang}, \binits{Z.}},
\oauthor{\bsnm{Sun}, \binits{L.}}:
Melodyglm: multi-task pre-training for symbolic melody generation.
arXiv preprint arXiv:2309.10738
(2023)
\end{botherref}
\endbibitem

\bibitem[\protect\citeauthoryear{Copet et~al.}{2024}]{copet2024simple}
\begin{botherref}
\oauthor{\bsnm{Copet}, \binits{J.}},
\oauthor{\bsnm{Kreuk}, \binits{F.}},
\oauthor{\bsnm{Gat}, \binits{I.}},
\oauthor{\bsnm{Remez}, \binits{T.}},
\oauthor{\bsnm{Kant}, \binits{D.}},
\oauthor{\bsnm{Synnaeve}, \binits{G.}},
\oauthor{\bsnm{Adi}, \binits{Y.}},
\oauthor{\bsnm{D{\'e}fossez}, \binits{A.}}:
Simple and controllable music generation.
Advances in Neural Information Processing Systems
\textbf{36}
(2024)
\end{botherref}
\endbibitem

\bibitem[\protect\citeauthoryear{Zhiqing et~al.}{2024}]{zhiqing2024text}
\begin{botherref}
\oauthor{\bsnm{Zhiqing}, \binits{H.}},
\oauthor{\bsnm{Rongjie}, \binits{H.}},
\oauthor{\bsnm{Xize}, \binits{C.}},
\oauthor{\bsnm{Yongqi}, \binits{W.}},
\oauthor{\bsnm{Ruiqi}, \binits{L.}},
\oauthor{\bsnm{Fuming}, \binits{Y.}},
\oauthor{\bsnm{Zhou}, \binits{Z.}},
\oauthor{\bsnm{Zhimeng}, \binits{Z.}}:
Text-to-song: Towards controllable music generation incorporating vocals and accompaniment.
arXiv preprint arXiv:2404.09313
(2024)
\end{botherref}
\endbibitem

\bibitem[\protect\citeauthoryear{Zeghidour et~al.}{2021}]{zeghidour2021soundstream}
\begin{barticle}
\bauthor{\bsnm{Zeghidour}, \binits{N.}},
\bauthor{\bsnm{Luebs}, \binits{A.}},
\bauthor{\bsnm{Omran}, \binits{A.}},
\bauthor{\bsnm{Skoglund}, \binits{J.}},
\bauthor{\bsnm{Tagliasacchi}, \binits{M.}}:
\batitle{Soundstream: An end-to-end neural audio codec}.
\bjtitle{IEEE/ACM Transactions on Audio, Speech, and Language Processing}
\bvolume{30},
\bfpage{495}--\blpage{507}
(\byear{2021})
\end{barticle}
\endbibitem

\bibitem[\protect\citeauthoryear{D{\'e}fossez et~al.}{2022}]{defossez2022high}
\begin{botherref}
\oauthor{\bsnm{D{\'e}fossez}, \binits{A.}},
\oauthor{\bsnm{Copet}, \binits{J.}},
\oauthor{\bsnm{Synnaeve}, \binits{G.}},
\oauthor{\bsnm{Adi}, \binits{Y.}}:
High fidelity neural audio compression.
arXiv preprint arXiv:2210.13438
(2022)
\end{botherref}
\endbibitem

\bibitem[\protect\citeauthoryear{Yu et~al.}{2024}]{yu2024suno}
\begin{botherref}
\oauthor{\bsnm{Yu}, \binits{J.}},
\oauthor{\bsnm{Wu}, \binits{S.}},
\oauthor{\bsnm{Lu}, \binits{G.}},
\oauthor{\bsnm{Li}, \binits{Z.}},
\oauthor{\bsnm{Zhou}, \binits{L.}},
\oauthor{\bsnm{Zhang}, \binits{K.}}:
Suno: potential, prospects, and trends.
Frontiers of Information Technology \& Electronic Engineering,
1--6
(2024)
\end{botherref}
\endbibitem

\bibitem[\protect\citeauthoryear{Amer et~al.}{2013}]{amer2013older}
\begin{barticle}
\bauthor{\bsnm{Amer}, \binits{T.}},
\bauthor{\bsnm{Kalender}, \binits{B.}},
\bauthor{\bsnm{Hasher}, \binits{L.}},
\bauthor{\bsnm{Trehub}, \binits{S.E.}},
\bauthor{\bsnm{Wong}, \binits{Y.}}:
\batitle{Do older professional musicians have cognitive advantages?}
\bjtitle{PloS one}
\bvolume{8}(\bissue{8}),
\bfpage{71630}
(\byear{2013})
\end{barticle}
\endbibitem

\bibitem[\protect\citeauthoryear{Mikutta et~al.}{2014}]{mikutta2014professional}
\begin{barticle}
\bauthor{\bsnm{Mikutta}, \binits{C.A.}},
\bauthor{\bsnm{Maissen}, \binits{G.}},
\bauthor{\bsnm{Altorfer}, \binits{A.}},
\bauthor{\bsnm{Strik}, \binits{W.}},
\bauthor{\bsnm{Koenig}, \binits{T.}}:
\batitle{Professional musicians listen differently to music}.
\bjtitle{Neuroscience}
\bvolume{268},
\bfpage{102}--\blpage{111}
(\byear{2014})
\end{barticle}
\endbibitem

\bibitem[\protect\citeauthoryear{Wang et~al.}{2024}]{wang2024muchin}
\begin{botherref}
\oauthor{\bsnm{Wang}, \binits{Z.}},
\oauthor{\bsnm{Li}, \binits{S.}},
\oauthor{\bsnm{Zhang}, \binits{T.}},
\oauthor{\bsnm{Wang}, \binits{Q.}},
\oauthor{\bsnm{Yu}, \binits{P.}},
\oauthor{\bsnm{Luo}, \binits{J.}},
\oauthor{\bsnm{Liu}, \binits{Y.}},
\oauthor{\bsnm{Xi}, \binits{M.}},
\oauthor{\bsnm{Zhang}, \binits{K.}}:
Muchin: A chinese colloquial description benchmark for evaluating language models in the field of music.
arXiv preprint arXiv:2402.09871
(2024)
\end{botherref}
\endbibitem

\bibitem[\protect\citeauthoryear{Dettmers et~al.}{2024}]{dettmers2024qlora}
\begin{botherref}
\oauthor{\bsnm{Dettmers}, \binits{T.}},
\oauthor{\bsnm{Pagnoni}, \binits{A.}},
\oauthor{\bsnm{Holtzman}, \binits{A.}},
\oauthor{\bsnm{Zettlemoyer}, \binits{L.}}:
Qlora: Efficient finetuning of quantized llms.
Advances in Neural Information Processing Systems
\textbf{36}
(2024)
\end{botherref}
\endbibitem

\bibitem[\protect\citeauthoryear{Bai et~al.}{2023}]{bai2023qwen}
\begin{botherref}
\oauthor{\bsnm{Bai}, \binits{J.}},
\oauthor{\bsnm{Bai}, \binits{S.}},
\oauthor{\bsnm{Chu}, \binits{Y.}},
\oauthor{\bsnm{Cui}, \binits{Z.}},
\oauthor{\bsnm{Dang}, \binits{K.}},
\oauthor{\bsnm{Deng}, \binits{X.}},
\oauthor{\bsnm{Fan}, \binits{Y.}},
\oauthor{\bsnm{Ge}, \binits{W.}},
\oauthor{\bsnm{Han}, \binits{Y.}},
\oauthor{\bsnm{Huang}, \binits{F.}}, et al.:
Qwen technical report.
arXiv preprint arXiv:2309.16609
(2023)
\end{botherref}
\endbibitem

\bibitem[\protect\citeauthoryear{Wu et~al.}{2023}]{wu2023large}
\begin{bchapter}
\bauthor{\bsnm{Wu}, \binits{Y.}},
\bauthor{\bsnm{Chen}, \binits{K.}},
\bauthor{\bsnm{Zhang}, \binits{T.}},
\bauthor{\bsnm{Hui}, \binits{Y.}},
\bauthor{\bsnm{Berg-Kirkpatrick}, \binits{T.}},
\bauthor{\bsnm{Dubnov}, \binits{S.}}:
\bctitle{Large-scale contrastive language-audio pretraining with feature fusion and keyword-to-caption augmentation}.
In: \bbtitle{ICASSP 2023-2023 IEEE International Conference on Acoustics, Speech and Signal Processing (ICASSP)},
pp. \bfpage{1}--\blpage{5}
(\byear{2023}).
\bcomment{IEEE}
\end{bchapter}
\endbibitem

\bibitem[\protect\citeauthoryear{Huang et~al.}{2022}]{huang2022mulan}
\begin{botherref}
\oauthor{\bsnm{Huang}, \binits{Q.}},
\oauthor{\bsnm{Jansen}, \binits{A.}},
\oauthor{\bsnm{Lee}, \binits{J.}},
\oauthor{\bsnm{Ganti}, \binits{R.}},
\oauthor{\bsnm{Li}, \binits{J.Y.}},
\oauthor{\bsnm{Ellis}, \binits{D.P.}}:
Mulan: A joint embedding of music audio and natural language.
arXiv preprint arXiv:2208.12415
(2022)
\end{botherref}
\endbibitem

\bibitem[\protect\citeauthoryear{Peebles and Xie}{2023}]{peebles2023scalable}
\begin{bchapter}
\bauthor{\bsnm{Peebles}, \binits{W.}},
\bauthor{\bsnm{Xie}, \binits{S.}}:
\bctitle{Scalable diffusion models with transformers}.
In: \bbtitle{Proceedings of the IEEE/CVF International Conference on Computer Vision},
pp. \bfpage{4195}--\blpage{4205}
(\byear{2023})
\end{bchapter}
\endbibitem

\bibitem[\protect\citeauthoryear{Ma et~al.}{2024}]{ma2024sit}
\begin{botherref}
\oauthor{\bsnm{Ma}, \binits{N.}},
\oauthor{\bsnm{Goldstein}, \binits{M.}},
\oauthor{\bsnm{Albergo}, \binits{M.S.}},
\oauthor{\bsnm{Boffi}, \binits{N.M.}},
\oauthor{\bsnm{Vanden-Eijnden}, \binits{E.}},
\oauthor{\bsnm{Xie}, \binits{S.}}:
Sit: Exploring flow and diffusion-based generative models with scalable interpolant transformers.
arXiv preprint arXiv:2401.08740
(2024)
\end{botherref}
\endbibitem

\bibitem[\protect\citeauthoryear{Kingma and Welling}{2013}]{kingma2013auto}
\begin{botherref}
\oauthor{\bsnm{Kingma}, \binits{D.P.}},
\oauthor{\bsnm{Welling}, \binits{M.}}:
Auto-encoding variational bayes.
arXiv preprint arXiv:1312.6114
(2013)
\end{botherref}
\endbibitem

\bibitem[\protect\citeauthoryear{Kong et~al.}{2020}]{kong2020hifi}
\begin{barticle}
\bauthor{\bsnm{Kong}, \binits{J.}},
\bauthor{\bsnm{Kim}, \binits{J.}},
\bauthor{\bsnm{Bae}, \binits{J.}}:
\batitle{Hifi-gan: Generative adversarial networks for efficient and high fidelity speech synthesis}.
\bjtitle{Advances in neural information processing systems}
\bvolume{33},
\bfpage{17022}--\blpage{17033}
(\byear{2020})
\end{barticle}
\endbibitem

\bibitem[\protect\citeauthoryear{Yang et~al.}{2023}]{yang2023baichuan}
\begin{botherref}
\oauthor{\bsnm{Yang}, \binits{A.}},
\oauthor{\bsnm{Xiao}, \binits{B.}},
\oauthor{\bsnm{Wang}, \binits{B.}},
\oauthor{\bsnm{Zhang}, \binits{B.}},
\oauthor{\bsnm{Bian}, \binits{C.}},
\oauthor{\bsnm{Yin}, \binits{C.}},
\oauthor{\bsnm{Lv}, \binits{C.}},
\oauthor{\bsnm{Pan}, \binits{D.}},
\oauthor{\bsnm{Wang}, \binits{D.}},
\oauthor{\bsnm{Yan}, \binits{D.}}, et al.:
Baichuan 2: Open large-scale language models.
arXiv preprint arXiv:2309.10305
(2023)
\end{botherref}
\endbibitem

\bibitem[\protect\citeauthoryear{Zeng et~al.}{2022}]{zeng2022glm}
\begin{botherref}
\oauthor{\bsnm{Zeng}, \binits{A.}},
\oauthor{\bsnm{Liu}, \binits{X.}},
\oauthor{\bsnm{Du}, \binits{Z.}},
\oauthor{\bsnm{Wang}, \binits{Z.}},
\oauthor{\bsnm{Lai}, \binits{H.}},
\oauthor{\bsnm{Ding}, \binits{M.}},
\oauthor{\bsnm{Yang}, \binits{Z.}},
\oauthor{\bsnm{Xu}, \binits{Y.}},
\oauthor{\bsnm{Zheng}, \binits{W.}},
\oauthor{\bsnm{Xia}, \binits{X.}}, et al.:
Glm-130b: An open bilingual pre-trained model.
arXiv preprint arXiv:2210.02414
(2022)
\end{botherref}
\endbibitem

\bibitem[\protect\citeauthoryear{Achiam et~al.}{2023}]{achiam2023gpt}
\begin{botherref}
\oauthor{\bsnm{Achiam}, \binits{J.}},
\oauthor{\bsnm{Adler}, \binits{S.}},
\oauthor{\bsnm{Agarwal}, \binits{S.}},
\oauthor{\bsnm{Ahmad}, \binits{L.}},
\oauthor{\bsnm{Akkaya}, \binits{I.}},
\oauthor{\bsnm{Aleman}, \binits{F.L.}},
\oauthor{\bsnm{Almeida}, \binits{D.}},
\oauthor{\bsnm{Altenschmidt}, \binits{J.}},
\oauthor{\bsnm{Altman}, \binits{S.}},
\oauthor{\bsnm{Anadkat}, \binits{S.}}, et al.:
Gpt-4 technical report.
arXiv preprint arXiv:2303.08774
(2023)
\end{botherref}
\endbibitem

\bibitem[\protect\citeauthoryear{Zhou et~al.}{2024}]{zhou2024lima}
\begin{botherref}
\oauthor{\bsnm{Zhou}, \binits{C.}},
\oauthor{\bsnm{Liu}, \binits{P.}},
\oauthor{\bsnm{Xu}, \binits{P.}},
\oauthor{\bsnm{Iyer}, \binits{S.}},
\oauthor{\bsnm{Sun}, \binits{J.}},
\oauthor{\bsnm{Mao}, \binits{Y.}},
\oauthor{\bsnm{Ma}, \binits{X.}},
\oauthor{\bsnm{Efrat}, \binits{A.}},
\oauthor{\bsnm{Yu}, \binits{P.}},
\oauthor{\bsnm{Yu}, \binits{L.}}, et al.:
Lima: Less is more for alignment.
Advances in Neural Information Processing Systems
\textbf{36}
(2024)
\end{botherref}
\endbibitem

\bibitem[\protect\citeauthoryear{Touvron et~al.}{2023}]{touvron2023llama}
\begin{botherref}
\oauthor{\bsnm{Touvron}, \binits{H.}},
\oauthor{\bsnm{Martin}, \binits{L.}},
\oauthor{\bsnm{Stone}, \binits{K.}},
\oauthor{\bsnm{Albert}, \binits{P.}},
\oauthor{\bsnm{Almahairi}, \binits{A.}},
\oauthor{\bsnm{Babaei}, \binits{Y.}},
\oauthor{\bsnm{Bashlykov}, \binits{N.}},
\oauthor{\bsnm{Batra}, \binits{S.}},
\oauthor{\bsnm{Bhargava}, \binits{P.}},
\oauthor{\bsnm{Bhosale}, \binits{S.}}, et al.:
Llama 2: Open foundation and fine-tuned chat models.
arXiv preprint arXiv:2307.09288
(2023)
\end{botherref}
\endbibitem

\bibitem[\protect\citeauthoryear{Ju et~al.}{2021}]{ju2021telemelody}
\begin{botherref}
\oauthor{\bsnm{Ju}, \binits{Z.}},
\oauthor{\bsnm{Lu}, \binits{P.}},
\oauthor{\bsnm{Tan}, \binits{X.}},
\oauthor{\bsnm{Wang}, \binits{R.}},
\oauthor{\bsnm{Zhang}, \binits{C.}},
\oauthor{\bsnm{Wu}, \binits{S.}},
\oauthor{\bsnm{Zhang}, \binits{K.}},
\oauthor{\bsnm{Li}, \binits{X.}},
\oauthor{\bsnm{Qin}, \binits{T.}},
\oauthor{\bsnm{Liu}, \binits{T.-Y.}}:
Telemelody: Lyric-to-melody generation with a template-based two-stage method.
arXiv preprint arXiv:2109.09617
(2021)
\end{botherref}
\endbibitem

\bibitem[\protect\citeauthoryear{Li et~al.}{2024}]{li2024scalability}
\begin{bchapter}
\bauthor{\bsnm{Li}, \binits{H.}},
\bauthor{\bsnm{Zou}, \binits{Y.}},
\bauthor{\bsnm{Wang}, \binits{Y.}},
\bauthor{\bsnm{Majumder}, \binits{O.}},
\bauthor{\bsnm{Xie}, \binits{Y.}},
\bauthor{\bsnm{Manmatha}, \binits{R.}},
\bauthor{\bsnm{Swaminathan}, \binits{A.}},
\bauthor{\bsnm{Tu}, \binits{Z.}},
\bauthor{\bsnm{Ermon}, \binits{S.}},
\bauthor{\bsnm{Soatto}, \binits{S.}}:
\bctitle{On the scalability of diffusion-based text-to-image generation}.
In: \bbtitle{Proceedings of the IEEE/CVF Conference on Computer Vision and Pattern Recognition},
pp. \bfpage{9400}--\blpage{9409}
(\byear{2024})
\end{bchapter}
\endbibitem

\bibitem[\protect\citeauthoryear{Liu et~al.}{2023}]{liu2023vit}
\begin{botherref}
\oauthor{\bsnm{Liu}, \binits{H.}},
\oauthor{\bsnm{Huang}, \binits{R.}},
\oauthor{\bsnm{Lin}, \binits{X.}},
\oauthor{\bsnm{Xu}, \binits{W.}},
\oauthor{\bsnm{Zheng}, \binits{M.}},
\oauthor{\bsnm{Chen}, \binits{H.}},
\oauthor{\bsnm{He}, \binits{J.}},
\oauthor{\bsnm{Zhao}, \binits{Z.}}:
Vit-tts: visual text-to-speech with scalable diffusion transformer.
arXiv preprint arXiv:2305.12708
(2023)
\end{botherref}
\endbibitem

\bibitem[\protect\citeauthoryear{Bao et~al.}{2022}]{bao2022all}
\begin{bchapter}
\bauthor{\bsnm{Bao}, \binits{F.}},
\bauthor{\bsnm{Li}, \binits{C.}},
\bauthor{\bsnm{Cao}, \binits{Y.}},
\bauthor{\bsnm{Zhu}, \binits{J.}}:
\bctitle{All are worth words: a vit backbone for score-based diffusion models}.
In: \bbtitle{NeurIPS 2022 Workshop on Score-Based Methods}
(\byear{2022})
\end{bchapter}
\endbibitem

\bibitem[\protect\citeauthoryear{Chen et~al.}{2024}]{chen2024pixart}
\begin{botherref}
\oauthor{\bsnm{Chen}, \binits{J.}},
\oauthor{\bsnm{Ge}, \binits{C.}},
\oauthor{\bsnm{Xie}, \binits{E.}},
\oauthor{\bsnm{Wu}, \binits{Y.}},
\oauthor{\bsnm{Yao}, \binits{L.}},
\oauthor{\bsnm{Ren}, \binits{X.}},
\oauthor{\bsnm{Wang}, \binits{Z.}},
\oauthor{\bsnm{Luo}, \binits{P.}},
\oauthor{\bsnm{Lu}, \binits{H.}},
\oauthor{\bsnm{Li}, \binits{Z.}}:
Pixart-$\backslash$sigma: Weak-to-strong training of diffusion transformer for 4k text-to-image generation.
arXiv preprint arXiv:2403.04692
(2024)
\end{botherref}
\endbibitem

\bibitem[\protect\citeauthoryear{Brooks et~al.}{2024}]{brooks2024video}
\begin{botherref}
\oauthor{\bsnm{Brooks}, \binits{T.}},
\oauthor{\bsnm{Peebles}, \binits{B.}},
\oauthor{\bsnm{Holmes}, \binits{C.}},
\oauthor{\bsnm{DePue}, \binits{W.}},
\oauthor{\bsnm{Guo}, \binits{Y.}},
\oauthor{\bsnm{Jing}, \binits{L.}},
\oauthor{\bsnm{Schnurr}, \binits{D.}},
\oauthor{\bsnm{Taylor}, \binits{J.}},
\oauthor{\bsnm{Luhman}, \binits{T.}},
\oauthor{\bsnm{Luhman}, \binits{E.}}, et al.:
Video generation models as world simulators.
https://openai. com/research/video-generation-models-as-world-simulators
(2024)
\end{botherref}
\endbibitem

\bibitem[\protect\citeauthoryear{Li et~al.}{2023}]{li2023mert}
\begin{botherref}
\oauthor{\bsnm{Li}, \binits{Y.}},
\oauthor{\bsnm{Yuan}, \binits{R.}},
\oauthor{\bsnm{Zhang}, \binits{G.}},
\oauthor{\bsnm{Ma}, \binits{Y.}},
\oauthor{\bsnm{Chen}, \binits{X.}},
\oauthor{\bsnm{Yin}, \binits{H.}},
\oauthor{\bsnm{Lin}, \binits{C.}},
\oauthor{\bsnm{Ragni}, \binits{A.}},
\oauthor{\bsnm{Benetos}, \binits{E.}},
\oauthor{\bsnm{Gyenge}, \binits{N.}},
\oauthor{\bsnm{Dannenberg}, \binits{R.}},
\oauthor{\bsnm{Liu}, \binits{R.}},
\oauthor{\bsnm{Chen}, \binits{W.}},
\oauthor{\bsnm{Xia}, \binits{G.}},
\oauthor{\bsnm{Shi}, \binits{Y.}},
\oauthor{\bsnm{Huang}, \binits{W.}},
\oauthor{\bsnm{Guo}, \binits{Y.}},
\oauthor{\bsnm{Fu}, \binits{J.}}:
MERT: Acoustic Music Understanding Model with Large-Scale Self-supervised Training
(2023)
\end{botherref}
\endbibitem

\bibitem[\protect\citeauthoryear{Rouard et~al.}{2023}]{rouard2022hybrid}
\begin{bchapter}
\bauthor{\bsnm{Rouard}, \binits{S.}},
\bauthor{\bsnm{Massa}, \binits{F.}},
\bauthor{\bsnm{D{\'e}fossez}, \binits{A.}}:
\bctitle{Hybrid transformers for music source separation}.
In: \bbtitle{ICASSP 23}
(\byear{2023})
\end{bchapter}
\endbibitem

\bibitem[\protect\citeauthoryear{D{\'e}fossez}{2021}]{defossez2021hybrid}
\begin{bchapter}
\bauthor{\bsnm{D{\'e}fossez}, \binits{A.}}:
\bctitle{Hybrid spectrogram and waveform source separation}.
In: \bbtitle{Proceedings of the ISMIR 2021 Workshop on Music Source Separation}
(\byear{2021})
\end{bchapter}
\endbibitem

\bibitem[\protect\citeauthoryear{McAuliffe et~al.}{2017}]{mfa}
\begin{bchapter}
\bauthor{\bsnm{McAuliffe}, \binits{M.}},
\bauthor{\bsnm{Socolof}, \binits{M.}},
\bauthor{\bsnm{Mihuc}, \binits{S.}},
\bauthor{\bsnm{Wagner}, \binits{M.}},
\bauthor{\bsnm{Sonderegger}, \binits{M.}}:
\bctitle{{Montreal Forced Aligner: Trainable Text-Speech Alignment Using Kaldi}}.
In: \bbtitle{Proc. Interspeech 2017},
pp. \bfpage{498}--\blpage{502}
(\byear{2017})
\end{bchapter}
\endbibitem

\bibitem[\protect\citeauthoryear{Donahue and Liang}{2021}]{donahue2021sheet}
\begin{botherref}
\oauthor{\bsnm{Donahue}, \binits{C.}},
\oauthor{\bsnm{Liang}, \binits{P.}}:
Sheet sage: Lead sheets from music audio.
Proc. ISMIR Late-Breaking and Demo
(2021)
\end{botherref}
\endbibitem

\bibitem[\protect\citeauthoryear{Dhariwal et~al.}{2020}]{dhariwal2020jukebox}
\begin{botherref}
\oauthor{\bsnm{Dhariwal}, \binits{P.}},
\oauthor{\bsnm{Jun}, \binits{H.}},
\oauthor{\bsnm{Payne}, \binits{C.}},
\oauthor{\bsnm{Kim}, \binits{J.W.}},
\oauthor{\bsnm{Radford}, \binits{A.}},
\oauthor{\bsnm{Sutskever}, \binits{I.}}:
Jukebox: A generative model for music.
arXiv preprint arXiv:2005.00341
(2020)
\end{botherref}
\endbibitem

\end{thebibliography}

\end{document}